\documentclass[twocolumn,prl]{revtex4}

\ifx\pdfoutput\undefined
  \usepackage[dvips]{graphicx}
\else
  \usepackage[pdftex]{graphicx}
\fi

\usepackage{dcolumn}
\usepackage{amsmath}
\usepackage{latexsym}

\begin{document}

\title[Short Title]{Rf-induced transport of Cooper pairs in superconducting single
electron transistors in a dissipative environment}

\author{S.~V.~Lotkhov}
\email[Electronic mail:~]{Sergei.Lotkhov@ptb.de}
\author{S.~A.~Bogoslovsky}
\author{A.~B.~Zorin}
\author{J.~Niemeyer}
\affiliation{Physikalisch-Technische Bundesanstalt, Bundesallee
100, 38116 Braunschweig, Germany}%

\date{\today}% It is always \today, today, but you may specify any date with \date.

\begin{abstract}
We investigate low-temperature and low-voltage-bias charge
transport in a superconducting Al single electron transistor in a
dissipating environment, realized as on-chip high-ohmic Cr
microstrips. In our samples with relatively large charging energy
values $E_{\text c} > E_{\text J}$, where $E_{\text J}$ is the
energy of the Josephson coupling, two transport mechanisms were
found to be dominating, both based on discrete tunneling of
individual Cooper pairs: Depending on the gate voltage $V_{\text
g}$, either sequential tunneling of pairs via the transistor
island (in the open state of the transistor around the points
$Q_{\text g} \equiv C_{\text g}V_{\text g} = e \bmod(2e)$, where
$C_{\text g}$ is the gate capacitance) or their cotunneling
through the transistor (for $Q_{\text g}$ away of these points)
was found to prevail in the net current. As the open states of our
transistors had been found to be unstable with respect to
quasiparticle poisoning, high-frequency gate cycling (at $f \sim
1$~MHz) was applied to study the sequential tunneling mechanism.
A simple model based on the master equation was found to be in a
good agreement with the experimental data.

\end{abstract}

\maketitle

\section{Introduction}
%\section{Introduction}

During the last few years, the ultrasmall Josephson junctions have
aroused the interest of researchers as convenient objects for
experimental investigations of macroscopic quantum effects. A
variety of their unique features formed the basis for possible
applications in quantum information processing (see, e.g., review
\cite{makhlin} and references therein), Cooper pair electrometry
\cite{Zorin-prl1}, and metrology (accurate current sources on the
basis of Cooper pair pumps \cite{Geerligs}).

The Cooper pair dynamics in such junctions appears as a result of
the competition between Coulomb and Josephson effects (see, e.g.,
\cite{LiZo}), and it is strongly affected by interaction with the
electrodynamic environment (for the review, see \cite{IngNaz}).
In particular, the Cooper-pair transport is very sensitive to
dissipative properties of the environment. Quantitatively, these
properties are characterized by the real part of the
frequency-dependent impedance of the environment Re$Z(\omega)$.

In the low-impedance case, $\left| Z(\omega) \right| \ll R_{\text
Q}$, where $R_{\text Q} \equiv h/4e^2 \approx 6.45~$k$\Omega$ is
the resistance quantum, the tunneling of Cooper pairs is elastic
and occurs at zero bias voltage as a coherent process. The
resulting net current presents the supercurrent as in the
"classical" Josephson junction \cite{Likh}, because the Josephson
phase across the junction is well defined. In most of the
experimental systems, small but finite dissipation of the
measuring circuitry $\left| Z(\omega) \right| \alt R_{\text Q}$
causes phase fluctuations which transform the zero voltage
supercurrent into a supercurrent peak of finite width located at
small voltages (see Ref.~\cite{Stein} and the references
therein). In contrast to the former two cases, tunneling of pairs
in the junction embedded in a rather dissipative environment,
${\text {Re}} Z(\omega) \agt R_{\text Q}$, is related to an
extensive energy exchange with the environment. Tunneling of
pairs in this case occurs incoherently and can be considered in
terms of individual events. The first observation of dissipative
tunneling of single Cooper pairs was reported by Kuzmin $et~al$.
\cite{Kuzmin} for a simple system of a small single junction
connected in series to a compact high-ohmic resistor.

\begin{figure}[t]
\begin{center}
\leavevmode
\includegraphics[width=3.2in]{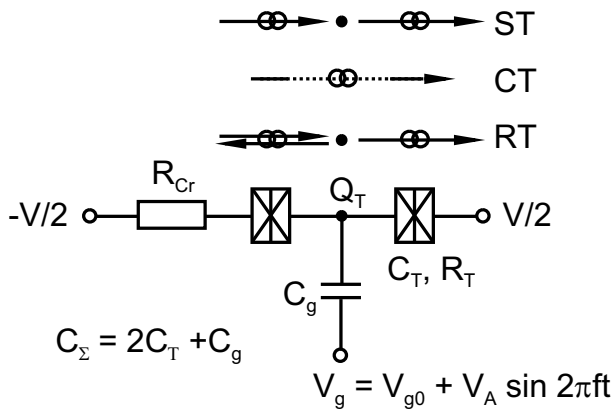}
\caption {Equivalent circuit diagram of a symmetric
superconducting SET and different mechanisms of Cooper pair
transport: sequential tunneling (ST), cotunneling (CT), and
resonant tunneling (RT).} \label{Equiv}
\end{center}
\end{figure}

Several different transport mechanisms were reported for a
two-junction superconducting device (single electron transistor,
SET) supplied by the high-ohmic resistor (see Fig.~\ref{Equiv}).
One of the mechanisms - sequential tunneling of pairs (ST) through
either junction - was studied theoretically by Wilhelm $et~al$.
\cite{Wilhelm}, and experimentally by Kycia $et~al$. \cite{Kycia}
and by Lu $et~al$. \cite{Lu}. These experiments were performed in
an Al SET positioned on top of a two-dimensional electron gas with
locally tunable dissipation (ground plane conductance). Another
mechanism based on a higher-order process of simultaneous
tunneling of pairs in both junctions - i.e., the cotunneling of
pairs (CT) - was recently studied in detail \cite{Lotk-prl2003}.
Moreover, resonant tunneling of single Cooper pairs (RT)
\cite{Brink} was found to be responsible for the charge transport
at somewhat larger bias voltages $\sim 2E_{\text c}/e <
2\Delta_{\text {Al}}$ \cite{Havi}. Here, $E_{\text c} \equiv
\frac{e^2}{2C_{\rm \Sigma}}$ is the charging energy, $C_{\rm
\Sigma} = 2C_{\text {T}} + C_{\text {g}}$ the total capacitance
of the transistor island, $C_{\text {T}}$ the tunnel capacitance
(in a symmetric device), $C_{\text {g}}$ the gate capacitance,
and $\Delta_{\text {Al}} \approx 200~\mu$eV the superconducting
energy gap of Al. The RT process involves resonant mixing of
charge states across one junction, completed by tunneling of
Cooper pairs in the second junction with energy dissipation in
the resistor.

In our recent paper \cite{Lotk-prl2003} we had shown that at low
temperatures and small voltage bias, CT dominates in the net
current for gate charges $Q_{\text g} \equiv C_{\text g}V_{\text
g}$, $\left| Q_{\text g} \right| \alt e/2$, where ST is blocked
due to the Coulomb barrier $\agt 2E_{\text c}$, provided the
Josephson coupling energy in the junctions $E_{\text J}$ is
small, $E_{\text J} \ll E_{\rm c}$. The ST process becomes
energetically favourable in the points around $Q_{\text g} \approx
e \bmod(2e)$, giving rise to the "open" states of the SET. In the
experiment, these states are often unstable towards spontaneous
tunneling of non-equilibrium quasiparticles (the so-called,
quasiparticle poisoning effect, see, e.g., \cite{Geerl} and
\cite{Tuom}). Study of the crossover between ST and CT is
therefore a challenging task.

In this paper, besides the observation of the CT current, we
succeeded in realizing of the regime, where the ST process
dominates over CT. We focus on the range of small bias voltages
$eV \ll 2\Delta_{\text {Al}}$, where tunneling of Cooper pairs
dominates over that of quasiparticles, and RT is energetically
unfavourable \cite{Brink}. In order to circumvent quasiparticle
poisoning of the island in the open states, we applied a
high-frequency signal of sufficient amplitude to the gate and
measured the time-averaged $I - V$ dependencies. The measured
curves were compared with the results provided by a simple model
based on the master equation approach. We considered the current
evolution along the gate cycle. As expected, it was characterized
by the major input from ST in the open states. Our calculation
also showed that the contribution of ST current, averaged over
the whole cycle, dominates over that of CT.

\section{Theoretical background}
%\section{Theoretical background}

At voltages $eV \ll 2\Delta_{\text {Al}} \approx 400~\mu$eV, which
are well below the thresholds for the different processes
incorporating tunneling of quasiparticles (see, e.g.,
Josephson-quasiparticle cycles in \cite{Brink}, \cite{Fult} and
\cite{TuoIEEE}), the tunneling of quasiparticles is rare
\cite{Tuom}, \cite{Amar}, and the net current is related to the
transport of Cooper pairs.

The Cooper pair dynamics in a superconducting SET with $E_{\text
J} \ll E_{\text c}$ and appreciable dissipation is qualitatively
similar to that of electrons in the normal-state SET described by
the orthodox theory \cite{Altsh}. In particular, a quantum state
of the whole system can be described by the number $n$ of extra
Cooper pairs on the island of the SET. The Josephson coupling in
the junctions can be regarded as a small perturbation in the total
Hamiltonian of the system, giving rise to random tunneling events
of different orders \cite{IngNaz}.

Since tunneling of Cooper pairs does not produce excitations, it
is dissipating properties of the environment that define the
functional dependence of a particular tunneling rate on the
electrostatic energy difference $E$ associated with the tunneling
event. The corresponding probability function $P(E)$
characterizes the ability of the environment to exchange an energy
$E$ with the tunneling pair. For a given temperature, the shape
of the $P$-function depends on the environmental impedance
$Z(\omega)$ as well as on the tunnel capacitances of the
junctions. In some important cases, $P(E)$ can be expressed
analytically (see, for example, below).

The process of the lowest order of perturbation is known as
direct tunneling of a pair in either junction. It changes the
charge state of the SET island by $ \pm 2e$. A rate of direct
tunneling can be expressed as \cite{AvNazOd}:
\begin{equation}
\label{Gamma} \Gamma (E)=\frac{\pi}{2\hbar} E_J^2 P(E).
\end{equation}
The validity range of the perturbative approach Eq.~(\ref{Gamma})
strongly depends on the dissipative properties of the environment
and is generally restricted to SETs with relatively weak
Josephson coupling and/or large damping, i.e., when:
\begin{equation}
\label{Valid} E_{\text J}P(E)\ll 1.
\end{equation}
An analysis performed in Ref.~\cite{IngNaz} shows that in the
case of zero temperature and a low-ohmic environment, Re$Z(\omega)
\ll R_{\text Q}$, the probability $P(E)$ grows infinitely around
zero energies, and the condition Eq.~(\ref{Valid}) is always
violated at small voltages. On the contrary, at finite
experimental temperatures $T \sim $~100~mK and/or in the case of
moderate dissipation, Re$Z(\omega) \sim R_{\text Q}$, the
function $P(E) \sim 4E_{\text c}^{-1}$ is distributed over the
broad range of energies so that the conditional
equation~(\ref{Valid}) can be reformulated as $\lambda \equiv
E_{\text J}/E_{\text c}\ll 4$. For the junctions in a high-ohmic
environment, Re$Z(\omega) > R_{\text Q}$, the conditional
equation~(\ref{Valid}) is even less restrictive at low energies,
where $P$ is a power function of $E$ (see Eq.~(111) in
Ref.~\cite{IngNaz}). As a result, in the systems with noticeable
dissipation, the approximation of discrete tunneling events can
even be applied to junctions with relatively strong Josephson
coupling.

For small energies, the external impedance can be considered to be
frequency-independent $Z(\omega) = Z(0)$ (purely ohmic) and one
obtains an analytic expression \cite{IngGrab} for the $P$-function
for tunneling in individual junctions of a symmetric transistor:
\begin{eqnarray}
\label{P_E} P(E) = &&\frac{z^{2z}} {8E_c } \left( \frac{2E_c}{\pi
^2 k_B T} \right)^{1 - 2z} e^{ - 2\gamma z} \nonumber \\
\times && \frac{\left| \Gamma \left[ z - j \left( E/2 \pi k_B T
\right) \right] \right|^2} {\Gamma \left( 2z \right)} \;
e^{E/2k_B T} ,
\end{eqnarray}
which is valid for low temperatures and energies $E \ll \hbar
\omega_Z$. The value of the threshold frequency $\omega_Z$ of the
ohmic approximation is defined for each experimental layout by
all environmental components (including resistor, tunnel and stray
capacitances and inductances according to the chip layout). In our
experiment, the typical value was evaluated $\omega_Z \sim 4
\times 10^{10}$~s$^{-1}$ ($\hbar \omega_Z \sim 25~\mu$eV, cf.
experimental section of this paper). In Eq.~(\ref{P_E}), $\Gamma$
is a gamma function and $\gamma\approx 0.577$ Euler's constant.
The dimensionless resistance $z = \frac{\rho}{4}$, where $\rho
\equiv Z(0)/R_{\text Q}$, accounts for the factor 1/4 (in
comparison to the case of a single junction or cotunneling) which
arises from the partial decoupling effect of the tunnel junction
from the environment due to the other junction (cf. network
considerations in Ref.~\cite{IngNaz}). The low-temperature
approximation requires $k_{\text B} T \ll 4E_{\text c}$/$\pi z$
which in most of the cases is valid in the whole mK temperature
range.

\begin{figure}[t]
\begin{center}
\leavevmode
\includegraphics[width=3.2in]{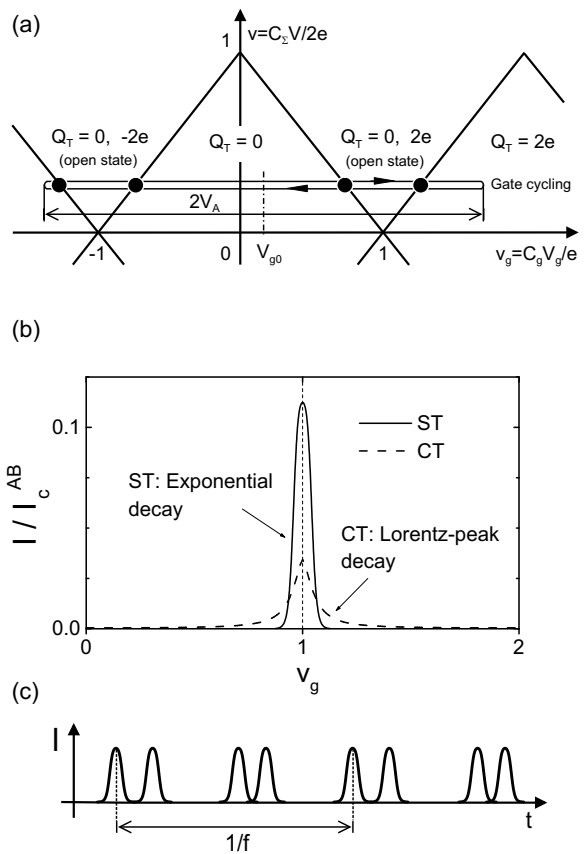}
\caption {(a) Stability diagram for Cooper pairs in a
single-charge transistor for $m = 0$. The closed circles show the
points of Cooper pair resonances in individual junctions, $E =
0$. The cycling amplitude $V_{\text A}$ is chosen to sweep over
two open domains. (b) ST and CT currents, normalized to the
Ambegaokar-Baratoff value $I_c^{\text {AB}}$, were calculated as a
function of the gate charge for the parameters of sample 2 (see
Table~\ref{Sampa}) and fixed $V = 25~\mu$V and $T = $~100~mK. (c)
Schematic representation of an rf-induced periodic sequence of
current pulses resulting from cycling over two open domains (as
shown in (a)). The amplitude of pulses given by the number of
tunneling pairs is fluctuating.}

\label{diag}
\end{center}
\end{figure}

For tunneling in the left (right) junction of the symmetric SET,
the energy difference is:
\begin{equation}
\label{Endif} E(V,Q) = eV - 4E_{\text c} \left( {1 \pm
\frac{Q}{e}} \right);
\end{equation}
it depends on voltage bias $V$ and effective charge $Q = Q_{\text
g}$ + $Q_{\text T}$, where $Q_{\text T} = 2ne + me$ is a charge
of $n$ extra Cooper pairs and $m$ extra quasiparticles on the
island. Similar to the case of a normal-state SET (see, e.~g.,
Ref.~\cite{Altsh}), the energy relations in Eq.~(\ref{Endif}) can
be mapped in the $V - V_{\text g}$ plane, giving a 2e-periodic
diamond-shaped stability diagram for Cooper pairs. As shown in
Fig.~\ref{diag}(a), this diagram is featured by a blockade domains
with the stable number of extra pairs on the island, as well as
by the regions of open states, where at least two alternating
charge states are available for realization of sequential
transport of Cooper pairs through the SET. At small biases, the
values of the gate charge for the open states of the SET, e.g., $
Q_{\text g} \approx e \bmod(2e)$ for $m = 0$, correspond to a
small vicinity of the resonance points $Q = \pm e$.

Similar to the normal-state SET, it is also possible to calculate
the net ST current at low temperatures, using the master equation
approach. In particular, in the important case of a small voltage
bias, the stationary population of the island is dominated by two
charge states, $n = 0,1$, with the occupation probabilities
$\sigma_{n} = \sigma_{n}(v,v_{\text g})$ which can be found with
the help of the detailed balance principle:
\begin{eqnarray}
\label{Popul} \sigma_{0} = \frac{\Gamma _1^- (1) +\Gamma _2^+
(1)}{\Gamma_{\Sigma}},
\nonumber \\
\sigma_{1} = \frac{\Gamma _1^+ (0) +\Gamma _2^-
(0)}{\Gamma_{\Sigma}},
\end{eqnarray}
where $\Gamma_{\Sigma} = \Gamma _1^- (1) + \Gamma _1^+ (0) +
\Gamma _2^+ (1) +\Gamma _2^- (0)$, and $\Gamma _{1,2}^{\pm} (n) =
\Gamma _{1,2}^{\pm}(v,v_{\text g},Q_{\text T})$ denote the rates
in the left (right) junction in the positive (negative) direction
and can be calculated with the help of Eqs.~(\ref{Gamma}),
(\ref{P_E}) and (\ref{Endif}). The net current is obtained as a
sum over the states $n$:
\begin{eqnarray}
\label{Master} I & = -2e \left[ \sigma_0 \Gamma _1^+ (0) -
\sigma_1 \Gamma _1^- (1) \right]
\nonumber \\
& = -2e \left[ \sigma_1 \Gamma _2^ + (1) - \sigma_0 \Gamma _2^ -
(0) \right].
\end{eqnarray}

An example of the ST current dependence on the gate voltages $I
(v,v_{\text g})$, calculated for fixed bias voltage and the
parameters of one of our samples, is shown in Fig.~\ref{diag}(b).
The net current, $I \sim 0.1I_{\text c}^{\text {AB}} \sim 1$~nA,
where $I_{\text c}^{\text {AB}}$ is the Ambegaokar-Baratoff
critical current, is significant in the open domain between the
resonant boundaries and decays exponentially with the height of
the barrier $\left| E \right|$, see Eq.~(\ref{Endif}), inside the
blockade domains: $I \propto {\text {exp}} (- \left| E
\right|/2k_{\text B} T)$.

Simple analysis of the time-dependent master equation shows that
the time $\tau$ the system takes to reach its stationary state,
when moving between the operating points in the open domain (where
the current is considerable), $\tau \sim \Gamma_{\Sigma}^{-1} \sim
e/2I \sim 0.1$~ns, is well below the experimental range of time
$\sim 10$~ns of passing one open domain when the gate is cycled at
$f \sim 1$~MHz. Making use of this fact, the quasi-stationary
master-equation approach can be applied and the stationary ST
currents considered for each point of the cycling trajectory, even
when modeling high-frequency behaviour of the system. The gate
cycling in particular will result in a regular pattern of ST
current pulses, for example, 4 pulses per cycle which involves 4
open domains (see Fig.~\ref{diag}(c)), and the mean ST current,
$\bar I_{\text {st}}$, can be obtained by averaging the values of
$I (v, v_{\text g})$ over the cycling period.

A contribution $I_{\text {ct}}$ to the total net current due to
higher-order processes, i.e., cotunneling of Cooper pairs, can be
analyzed using the lines developed in Ref.~\cite{Lotk-prl2003}.
It was shown in particular that an SET can be modeled by a single
junction with an effective, gate-voltage-dependent, Josephson
coupling Hamiltonian  $H_{\text {tr}}(\phi,Q)$, where $\phi$ is
the overall Josephson phase. The coupling is $2\pi$-periodic in
$\phi$, but its shape is generally different from $-{\text
{cos}}\phi$. Expansion of the coupling Hamiltonian into Fourier
series (see Eq.~(4) in Ref.~\cite{Lotk-prl2003}) makes it
possible to assess the individual contributions to the CT-current
from the different-order cotunneling channels, each one
simultaneously transporting a certain number $k$ of Cooper pairs
through the SET. Similar to direct tunneling, the rate formula in
Eq.~(\ref{Gamma}) can be used to calculate the partial
cotunneling rates due to each channel, with the substitution $E =
2keV$, taking account of cotunneling of $k$ pairs. The effective
value of Josephson coupling energy for each channel is given by
the corresponding Fourier coefficient $E^{(k)}_{\text J}(Q)$, and
the dimensionless resistance $k^2 \rho$ must be used in
Eq.~(\ref{P_E}) for the $P$-function.

It was shown in Ref.~\cite{Lotk-prl2003} that the lowest-order,
i.e. single-pair, CT-process with $k = 1$  generally dominates in
the CT-current through the SET, and the principal Fourier term can
provide a reasonable approximation for the coupling Hamiltonian:
$H_{\text {tr}}(\phi,Q) \approx -E^1_{\text J}(Q)$cos($\phi$). It
can further be demonstrated that dropping-out of the higher terms
causes an error of $\delta I_{\text {ct}} \sim 10$~\% at most
which is obtained, in a symmetric SET, in the degeneracy points
$Q \approx \pm e$ (see Eq.~6 in Ref.~\cite{Lotk-prl2003}). Around
these points, the CT-current almost behaves like a Lorentzian
peak, and at $Q \to 0$ it decays more slowly than the
exponentially suppressed ST current (see Fig.~\ref{diag}(b)). As
a result, the CT-current dominates in the net current inside the
blockade domains.

\section{Experimental setup}
%\section{Experimental setup}

\begingroup
\squeezetable
\begin{table*}[t]
\caption{\label{Sampa} Sample parameters.}
\begin{ruledtabular}
\begin{tabular}{l @{\hfill \vline} ccccccc}
Sample~~~ &$R_{\text T} (k\Omega)$ & $R_{\text {Cr}} (k\Omega)$ &
$I_{\text c}^{\text {AB}}$ (nA) & $E_{\text J}$ ($\mu$eV) &
$E_{\text c}$ ($\mu$eV) & $E_{\text J}/E_{\text c}$\\
\hline
  ~~~~1 & 20 & 0 & 15 & 32 & 150 & 0.22\\
  ~~~~2 & 20 & 3 & 15 & 32 & 150 & 0.22\\
  ~~~~3 & 25 & 12 & 13 & 26 & 170 & 0.15\\
\end{tabular}
\end{ruledtabular}
\end{table*}
\endgroup

\begin{figure}[t]
\begin{center}
\leavevmode
\includegraphics[width=3.2in]{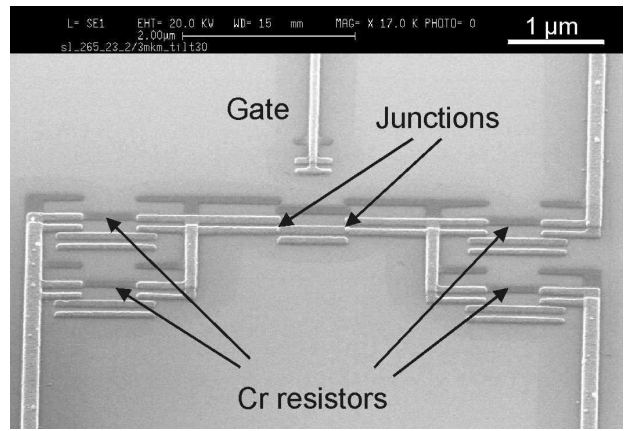}
\caption {SEM-micrograph of the Al/AlO$_{\text {x}}$/Al-Cr -
transistor. The configuration with two parallel Cr microstrips on
each side was implemented for the purpose of directly measuring of
$R_{\text {Cr}}$ as half the resistance between both right-hand
arms of the device.} \label{SEM}
\end{center}
\end{figure}

The samples were fabricated by the shadow evaporation technique.
The details of our fabrication process can be found in
\cite{Napoli}. Each structure was built on three almost congruent
metal layers (see Fig.~\ref{SEM}) deposited at different angles
through the same pattern in the mask with hanging bridges. The
layers were of Cr, 7~nm thick and 0.1~$\mu$m wide as used for
high-ohmic resistors, and two layers of Al, 30 and 35~nm in
thickness, evaporated with intermediate oxidation of the first
layer to form the tunnel barriers. The nominal tunnel junction
area was 40~nm $\times$ 80~nm.

The results reported in this paper were obtained on two SETs
(samples 2 and 3, see Table~\ref{Sampa}) in a dissipating
environment, realized as Cr microresistors with $Z(0) = R_{\text
{Cr}} \approx $~3 and 12~k$\Omega$ as well as on a reference
sample 1 without resistor, i.e., $R_{\text {Cr}} = $~0. The
corresponding values of the dimensionless resistance parameter
were $\rho \approx $~0.5,~2 and 0. Due to the linearity and small
lengths of the resistors it was possible to consider them as
lumped elements (ohmic resistors) and neglect their distributed
self-capacitance ($\approx$ 65~aF/$\mu$m, as earlier estimated
for the similar layout \cite{Zang}).

Ambegaokar-Baratoff critical currents of individual junctions
were estimated, assuming symmetry of the transistors: $I_c^{\text
{AB}} = \frac {\pi \Delta_{\text {Al}}}{2e R_{\text {T}}} \approx
$ 13---15~nA, where $R_{\text {T}}$ is the tunnel resistance. The
corresponding values of the Josephson energy were $E_{\text J}=
\frac{\Phi_{\text 0}} {2\pi}I_{\text c}^{\text {AB}} \approx $
26---32~$\mu$eV, where $\Phi_{\text 0}=\frac{h}{2e}$ is the flux
quantum.  Roughly evaluated with the help of a similar device
with $R_{\text {Cr}} = 0$ in the normal state, the characteristic
charging energy of the transistor island was $E_{\rm c} \approx$
150 to 170~$\mu$eV. The ratio of the characteristic energies in
all our samples, $\lambda \approx $ 0.15---0.2, indicates that the
SET charging energy dominated over the Josephson energy of the
junctions. 1e-modulation period in the $I - V_{\text g}$ curve
without rf drive was $\approx 20$~mV yielding $C_{\text g}
\approx 8$~aF.

$I - V$ characteristics were measured at the base temperature of
the dilution refrigerator of about 10~mK. The bias and gate lines
were equipped with microwave filters made of
Thermocoax$^{\rm\textregistered}$ cables, 1.5m in length, each
thermally anchored to the mixing chamber. The filters provided
reliable electromagnetic isolation of the sample inside the
shielded sample holder at frequencies above $f \sim 1$~GHz
\cite{Zorin-Thermocoax}.

\section{Experimental results}
%\section{Experimental results}

As discussed above, observation of ST current should be possible
in the vicinity of the resonance points $Q = \pm e$. The parity
effect in a superconducting SET (see, e.~g., Refs.~\cite{Tuom},
\cite{Amar} and \cite{Parity}), basically makes it possible to
reach these points through appropriately setting the gate voltage,
provided quasiparticle poisoning (QPP) effects are absent (the
value of $m$ is zero or, at least, fixed), and the experimental
gate modulation is 2e-periodical. Unfortunately, the effect of
QPP did not allow us to reach the resonance points and hindered
the observation of ST currents in the dc regime: The observed
periodicity was strictly 1e, corresponding to the range of the
island charge $-e/2 < Q < e/2$, i.~e., inside the ST blockade
domain at low bias voltages.

The effect of QPP is generally supposed to originate from various
experimental non-idealities including subgap quasiparticle states
\cite{Lafa}, out-of-equilibrium quasiparticles \cite{Joyez}, back
influence of the SET electrometer on the island (box)
\cite{Mann}, etc. It involves uncontrolled excitation of
quasiparticles in the outer electrodes and their tunneling into
the island of SET, thus changing $m$ and minimizing the charging
energy of the system. In particular, in the resonant points $
Q_{\text g} = e \bmod(2e)$ and $V = 0$, the charging energy is
minimized, when one extra quasiparticle enters the island
resulting in $Q = 0$ and Coulomb blockade of ST. The poisoning
rate $\gamma_{\text B}$ depends on the number of excitations and
the subgap tunnel resistance. Typically $\gamma_{\text B} \sim $
16---1000~s$^{-1}$ (see, e.g., Ref.~\cite{Mann}).

\begin{figure}[t]
\begin{center}
\leavevmode
\includegraphics[width=3.2in]{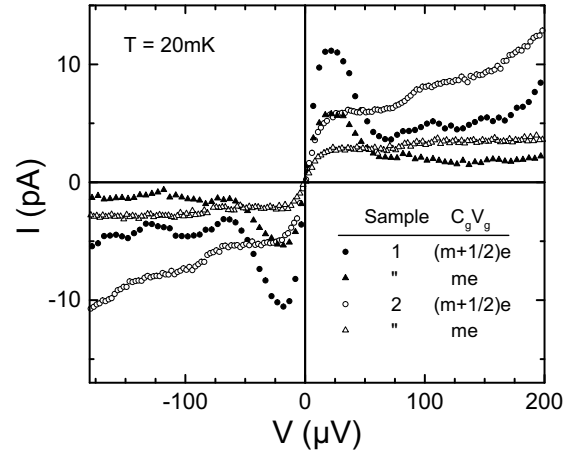}
\caption {Dc $I - V$ curves of two SETs with and without on-chip
resistor, samples 1 and 2, demonstrate the effects of both gate
voltage and dissipative suppression of cotunneling. } \label{dc}
\end{center}
\end{figure}

In Fig.~\ref{dc} the dc $I - V$ curves are shown for integer and
half-integer values of $Q$ in the system with considerable
dissipation ($R_{\text {Cr}} = 3$~k$\Omega$, sample 2) in
comparison to those for a similar SET fabricated without
engineered dissipation ($R_{\text {Cr}} = 0$, sample 1). As
expected for the blockade state, the dc currents were small (up to
$\sim 10$~pA) and could be attributed to the higher-order CT
process. In good agreement with the theory \cite{Lotk-prl2003},
the CT-current was modulated by the gate with a
maximum-to-minimum ratio of $\approx$2 within the accessible
range of $-e/2 < Q < e/2$.

To avoid the effect of QPP and reach the open domain for ST at
least for a short time, we applied a periodic gate signal of
frequency $f \sim 1$~MHz which was well above a typical
quasiparticle tunneling rate. Due to these fast oscillations, the
actual values of $Q$ along the gate sweep were expected to be
less affected by the relatively rare quasiparticle tunneling than
in the dc mode. As a result, the net current averaged over many
cycles should include a substantial contribution due to ST in the
open domains.

\begin{figure}[t]
\begin{center}
\leavevmode
\includegraphics[width=3.2in]{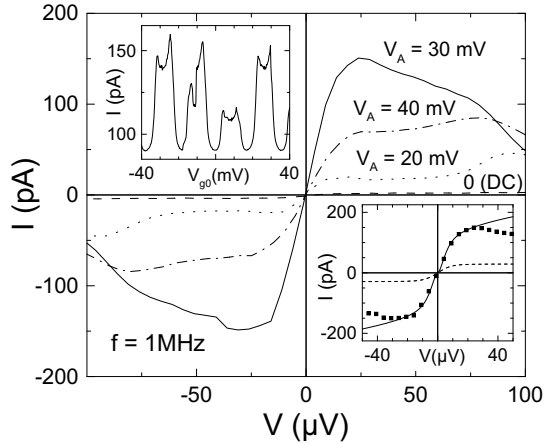}
\caption {Rf-induced currents in sample 2 ($R_{\text {Cr}}
\approx 3~k\Omega$) for different cycling amplitudes. The cycle
was centered in the middle of the blockade domain. $Top~inset$:
Rf-induced current vs. dc gate shift $V_{g0}$ measured at $V =
50~\mu$V for cycling at $f = 1$~MHz and a relatively small
amplitude $V_{\text A} = 20$~mV, i.~e., $C_{\text g}V_{\text A}/e
= 1$. $Bottom~inset$: $I - V$ curve (symbols) at $V_{\text A}$ =
30~mV ($C_{\text g}V_{\text A}/e = 1.5$) compared with its fit
$\bar I_{\text {tot}}(V)$ (solid line) made at fixed $E_{\text c}
= 150~\mu$eV. The fitting values were: $E_{\text J} = 40~\mu$V,
$R_{\text {Cr}} = 3.3~k\Omega$, $C_{\text g}V_{\text A}/e = 1.3$,
$T = $100 mK. The dashed line shows the calculated contribution
of CT, $\bar I_{\text {ct}}(V)$.}

\label{rf-Rcr-small}
\end{center}
\end{figure}

The $I-V$ curves measured in this regime are shown in
Fig.~\ref{rf-Rcr-small} for sample 2 and several rf amplitudes of
the gate signal $V_{\text A}$, $1 < C_{\text g}V_{\text A}/e < 2$.
The most striking feature of these curves is the dramatic
increase in the SET current, demonstrating a large contribution
from the open domains involved in gate cycling (cf. scale of
rf-induced current in Fig.~\ref{rf-Rcr-small} with that of dc $I
- V$ curves of this sample in Fig.~\ref{dc}). At fixed bias, the
rf-induced current was a 1e-periodic function of the dc gate shift
$V_{g0}$ (cf. top inset in Fig.~\ref{rf-Rcr-small}) which shows
that the effect of QPP was still substantial in a long time scale.

A rough evaluation of the contribution of both ST and CT to the
average current in the rf mode can be made with the help of
duty-cycle weighting of typical open domain values. For sample 2,
these values were estimated to be $I_{\text {st}} \approx 1.5$~nA
and $I_{\text {ct}} \approx 0.3$~nA, respectively. Further
assuming $V = 25~\mu$V and $V_{\text A}$ = 30~mV, the duty cycle
is $\approx 0.05$  producing the estimates: $\bar I_{\text {st}}
\approx $ 100~pA and $\bar I_{\text {ct}} \approx $ 20~pA,
respectively. The total current value $\bar I_{\text {tot}} =
\bar I_{\text {st}} + \bar I_{\text {ct}} \approx $ 120~pA thus
obtained agree with the order of magnitude of the experimental
data in Fig.~\ref{rf-Rcr-small}.

For more accurate evaluation of $\bar I_{\text {st}}$, we applied
the quasi-static master equation approach
(Eqs.~(\ref{Popul})-(\ref{Master})) and, for the total current
$\bar I_{\text {tot}}$, also took the cycle average $\bar
I_{\text {ct}}$ of CT current into account. We found satisfactory
agreement between calculated and experimental $I - V$ curves in
the low voltage region, $V \alt 25~\mu$V (see the bottom inset in
Fig.~ \ref{rf-Rcr-small}). Noticeable discrepancy of the data at
higher voltages in our opinion points to the upper limit of the
frequency-independent behaviour of the environmental impedance
assumed in Eq.~(\ref{P_E}) for the calculation of the
$P$-function. It is still difficult to explain the relatively low
value of the cutoff frequency $\omega_Z = eV/\hbar \sim 4 \times
10^{10}$~s$^{-1}$ for our short ($\sim 0.5~\mu$m - long)
microstrips (cf. cut-off frequencies $ \sim 10^{11}$~s$^{-1}$ for
Re$Z(\omega)$ of similar microstrips in Ref.~\cite{Zang}).

\begin{figure}[t]
\begin{center}
\leavevmode
\includegraphics[width=3.2in]{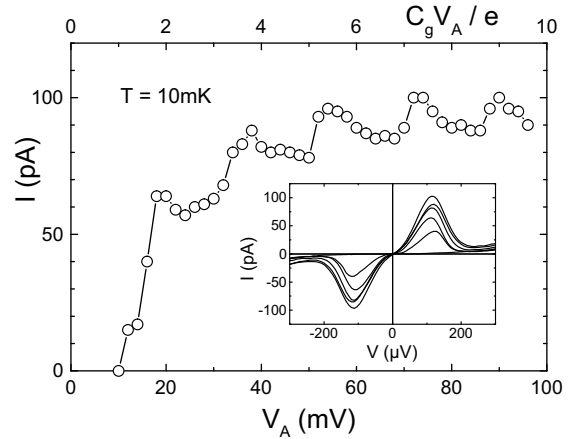}
\caption {Dependence of rf-induced current peak heights on the
amplitude of rf cycling in sample 3 with significant dissipation
($R_{\text {Cr}} \approx 12~k\Omega$). The gate cycling frequency
was $f$ = 3~MHz. The cycle was centered in the middle of the
blockade domain. The shapes of rf-induced current peaks for this
device are presented in the inset for different amplitudes
$V_{\text A}$ = 12, 16, 20, 40, 60, 72~mV (in the order of peak
increase). Note the significant blockade at the smallest cycling
amplitude $V_{\text A} = 12$ caused by insufficient gate sweep
and small value of CT-current $I_{\text {ct}} \sim 1$~pA at $V
\sim $100---200$\mu$eV. }

\label{rf-Rcr-large}
\end{center}
\end{figure}

The height of the current peaks was found to quasi-periodically
depend on the drive amplitude, with a tendency to saturate at
higher amplitudes (cf. Fig.~\ref{rf-Rcr-large}, showing the
dependence for sample 3 with $R_{\text {Cr}} \approx
12~k\Omega$). In our opinion, the observed quasi-periodicity on
$V_{\text A}$ is a result of the step-wise involvement of multiple
open domains in gate cycling, each contributing one pulse of ST
current. In good agreement with this model, every period consists
of two distinct segments: sharp increase in the peak height just
above the values $V_{\text A} = (2n+1)e/C_{\text g}$, $n
=$~0,1,2,..., where an additional open domain becomes involved,
and gradual reduction of the height along the other part of the
period, corresponding to a shortening of the duty cycle with
increasing amplitudes. The inset in Fig.~\ref{rf-Rcr-large} shows
the rf-induced current peaks as a function of transport voltage
$V$. The position of the peaks was evidently shifted to higher
voltages as compared to the sample 2, which results from the
shape of the $P$-function for a relatively large dissipation in
sample 3.

\section{Conclusions}
We reported the experimental results on dissipative sequential
tunneling of Cooper pairs in SETs with compact on-chip resistors.
Whereas Cooper pair cotunneling was the dominant dc transport
mechanism for the gate voltage range $-e/2 < C_{\text g}V_{\text
g} < e/2$, we were able to observe rf-induced sequential
tunneling of pairs realized with the help of high-frequency gate
cycling resulted in a pulsed ST current through the SET. We
demonstrated that - in accordance with theoretical predictions -
the integral contribution to the net current of ST exceeds that of
the CT in the rf-cycling mode, indicating its dominance in the
vicinity of the resonant points $Q = \pm e$. By a simple model
based on a quasi-static master equation approach it was possible
to well describe the observed $I-V$ curve at low voltages.

Moreover, we demonstrated that, due to the effectively smaller
environmental resistance seen by each single junction of an SET
($\sim \frac{1}{4} R_{\text {Cr}}$), the appreciable rf-induced ST
currents can be realized even in systems with larger resistors of
several $R_{\text Q}$, where the CT current is substantially
suppressed. This effect, for example, is important for possible
application of discrete Cooper pair tunneling to generate
accurate currents by Cooper-pair pumps \cite{Geerligs}.

Another important result of our studies is the demonstration of
quasiparticle-poisoning-free performance of a superconducting SET
at frequencies of $f \sim $ 1~MHz. This is important for the
potential application of small Josephson junctions in quantum
information processing, giving an estimate for a period of time
of at least $\sim 1~\mu$s \cite{KHz} as is available for reading
out the final state of a charge qubit \cite{Qubit} of the loop
configuration with nonzero circulating supercurrent.

\section*{ACKNOWLEDGMENTS}
The work was partially supported by the EU through projects COUNT
and SQUBIT-2.

%\section*{REFERENCES}


\begin{references}
\bibitem{KHz} In some experiments, we were able to trace gradually decreasing rf-induced currents
down to frequencies of $\sim~1$~kHz, which indicates that even
longer times are, in principle, available for QPP-free operation.
\end{references}
\end{document}